\documentclass[twocolumn,showpacs,preprintnumbers,amsmath,amssymb]{revtex4}
\usepackage{graphicx}
\begin{document}
\title{ Magnetoplasmons in  layered graphene structures  }
\author{  Oleg L. Berman,$^{1}$ Godfrey Gumbs$^{2}$,
and Yurii E. Lozovik,$^{3}$}
\affiliation{\mbox{$^{1}$Physics Department, New York City
College of Technology of the
City University of New York,} \\ 300 Jay Street, Brooklyn, NY 11201 \\
\mbox{$^{2}$
Department of Physics and Astronomy,
Hunter College of the City University of New York,} \\
695 Park Avenue, New York, NY 10021 \\
\mbox{$^{3}$Institute of Spectroscopy, Russian Academy of
Sciences, 142190 Troitsk, Moscow Region, Russia}}

\date{\today}

\begin{abstract}
We calculate the dispersion equations for magnetoplasmons  in a
single layer,   a pair of parallel layers, a graphite bilayer and a
superlattice of graphene layers in a  perpendicular  magnetic field.
We demonstrate the feasibility of a drift-induced instability of
magnetoplasmons  The magnetoplasmon instability
in a  superlattice is enhanced compared to a single graphene layer.
The energies of the unstable magnetoplasmons  could be in the terahertz (THz) part of the
electromagnetic spectrum. The enhanced instability makes superlattice graphene a potential
source of THz radiation.

\end{abstract}


\maketitle

\section{Introduction}
\label{intro}

Recent advances in fabrication techniques  have made it possible to
produce graphene, which is a two-dimensional (2D)  honeycomb lattice
of carbon atoms forming the basic planar structure in graphite
\cite{Novoselov1}. Graphene has stimulated considerable theoretical
interest as a semi-metal  whose electron effective mass may be
described by an unusual massless Dirac-fermion band structure.
Several novel many-body effects in graphene have been investigated
\cite{Shung1,DasSarma}.  The theory of Weiss oscillations in the
magnetoplasmon spectrum of Dirac electrons in graphene has been
developed in Ref.~[\onlinecite{nnew1}]. The magnetoplasmon
excitations in graphene for filling factors $\nu < 6$ has been
calculated in Ref.~[\onlinecite{nnew2}]. In recent experiments, the
integer quantum Hall effect (IQHE) has been reported
\cite{Novoselov2}. Quantum Hall ferromagnetism in graphene has been
investigated from a theoretical point of view \cite{Nomura}.
Graphene has a number of interesting properties as a result of its
unusual band structure which is linear near two inequivalent points
($K$ and $K^\prime$) in the Brillouin zone. In the presence of a
magnetic field, the graphene structure shifts both the Shubnikov–de
Haas oscillations \cite{Mikitik} as well as
 the step pattern of the IQHE \cite{Ando}. Both these effects have recently
been reported experimentally \cite{Novoselov2}. The spectrum
of plasmon excitations in a single graphene layer embedded  in a material
with effective dielectric constant $\varepsilon_b$ in the absence of an external
magnetic field, has been calculated in \cite{Hwang_DasSarma}. In this
paper, we show that features, such as charge density oscillations, arise
when a magnetic field is applied.

This paper is organized as follows. In Sec.\ \ref{single}, we
analyze  the magnetoplasmon spectrum for a single graphene layer.
The collective  charge density excitations in a strong magnetic
field for a graphite bilayer and a bilayer graphene are calculated
in Secs.\ \ref{bila} and \ref{bilag}, respectively.  The enhancement
of the magnetoplasmon instability in an infinite periodic graphene
superlattice is investigated  in Sec.\ \ref{inf}. In these
calculations, we assume that there is no tunneling between the
graphene layers forming the superlattice. The results of our
numerical calculations are presented in for each structure
investigated. A brief discussion of plasmon instabilities in
graphene is presented in Sec.\ \ref{disk}.

\section{A Single graphene layer}
\label{single}

Let us first consider electrons in a single graphene layer in the $xy$-plane
in a perpendicular magnetic field  $\mathbf{B}$ parallel to the
positive $z$ axis. Here, we neglect the Zeeman splitting
and assume valley energy degeneracy, describing the eigenstates
by two pseudo spins \cite{Ando,Jain}. We have an effective $2\times 2$ matrix
Hamiltonian $\hat{H}_{(0)}$ whose diagonal elements are zero and whose off-diagonal elements are
$\hat{\pi}_x \pm  i \hat{\pi}_y$
where $\hat{\mathbf{\pi}} = -i\hbar\nabla +e\mathbf{A}$, $-e$
is the electron charge, $\mathbf{A}$ is the vector potential,
 $v_F = \sqrt{3}at/(2\hbar)$ is the Fermi velocity with $a =2.566 \AA$
denoting the lattice constant, and $t \approx 2.71$ eV is the overlap integral
between  nearest-neighbor carbon atoms \cite{Ando}.

Choosing $\mathbf{A} = (0,Bx,0)$, the eigenfunctions
of $\hat{H}_{(0)}$ are labeled by $\displaystyle{\alpha=\{k_y,n,s(n)\}}$,
where $n=0,1, 2,\cdots$ is
the Landau level index, $k_y$ is the electron wave vector in the
$y$-direction, and $s(n)$, which is defined by
$s(n) =0$ for $n=0$ and $s(n) =\pm 1$ for $n>0$,
labels the conduction ($+1$) and valence ($-1$ and $0$) band,
respectively. The eigenfunction  $\psi_\alpha(\mathbf{r})$
is given  by a spinor $\psi_\alpha(x,y)$ with components given by \cite{Ando}
$\psi_\alpha^{(1)}=C_{n}  e^{ik_y y}s(n) i^{n-1}\Phi_{n-1}(x+l_H ^{2}k_y )/ \sqrt{L_y} $ and
$\psi_\alpha^{(2)}=C_{n}  e^{ik_y y}i^{n}\Phi_{n}(x+l_H ^{2}k_y )/ \sqrt{L_y}$. Here,
$l_H = \sqrt{\hbar /eB}$, and $L_y$ is a normalization length.
We have $C_n =1$ for $n=0$,  $C_n =1/\sqrt 2$ for $n>0$ and
$\Phi_n(x) = \left(2^{n}n!\sqrt{\pi}l_H \right)^{-1/2}
e^{-(x/l_H )^2/2} H_n\left(x/l_H \right)$, where $H_n(x)$ is a
Hermite polynomial. The eigenenergies are given by
$\displaystyle{\epsilon_\alpha = s(n) \epsilon_{n} = s(n) (\hbar v_F /l_H )
\sqrt{2n}}$, for which successive levels are not equally separated.

The dynamic dielectric function  in RPA \cite{Pines} is given by
$\varepsilon(q,\omega)= 1 - V_{c}(q)\Pi(q,\omega)$, where $q$ is the
in-plane wave vector,  $V_{c}(q) =  2\pi e^2/(\varepsilon_s q)$
is the 2D Coulomb interaction
 and the 2D polarization function is

\begin{eqnarray}
 & &\Pi(q,\omega)
\nonumber\\
&=& \frac{g_{s}g_{v}}{2\pi l_H ^{2}}
\sum_{n=0}^\infty\sum_{n^\prime=0}^\infty \sum_{s(n),s^\prime (n^\prime)}
\frac{f_{s(n)n}
-f_{s^\prime(n^\prime)n^\prime}}
{\hbar\omega+\epsilon_{s(n)n}-\epsilon_{s^\prime(n^\prime)n^\prime}} \nonumber \\
&\times & F_{s(n)s^\prime(n^\prime)}(n,n^\prime,q)\ ,
\label{pol}
\end{eqnarray}
where  $f_{s(n)n}$ is the Fermi-Dirac  function,
$F_{ss^\prime}(n,n^\prime)$  arises from
the overlap of eigenstates and is given by

\begin{eqnarray}
&& F_{ss^\prime }(n,n^\prime ,q)=
C_{n_{1}}^{2}C_{n_{2}}^{2}
\left[-\frac{q^{2}l_H ^{2}}{2}\right]^{n_{1}-n_{2}}
\frac{1}{\left|(n_{1}-n_{2})!\right|^{2}}
\nonumber \\ &\times&  \left(
s_{1}(n_{1})s_{2}(n_{2})
\left|\frac{(n_{1}-1)!}{(n_{2}-1)!}\right| +
\left|\frac{n_{1}!}{n_{2}!}\right|\right)\ .
\label{Fkk13}
\end{eqnarray}

The magnetoplasmon  dispersion relation for a single graphene layer
was obtained by seeking the solutions of $\varepsilon(q,\omega) = 0$.
The     highest valence band is full and all others empty at $T=0$ K.
Transitions to the lowest five Landau
levels in the conduction and valence bands were the only single-particle excitations
included in our calculations.
Fig.\  \ref{fsingle} (a) is the solution of the dispersion
equation for a single layer of graphene when the imaginary part
of the plasmon frequency is zero. In this case, the plasmons are
self-sustaining oscillations except when they enter the particle-hole mode
region where they undergo a loss due to Landau damping. In Fig.\  \ref{fsingle} (b),
we plot the solutions of the dispersion when the frequency is
complex for which the real part that is linear in $q$
and exists only where the   magnetoplasmon in Fig.\  \ref{fsingle} (a)
has negative group velocity.   The real and imaginary parts of the frequency are denoted by
$\omega_R$ and $\omega_I>0$, respectively.
The loss corresponds to finite imaginary part of frequency.
This instability after excitation
is due to a transfer of energy back
from a magnetoplasmon to an electric current which excites it,
thereby making this collective mode unstable (see Fig\  \ref{fsingle} (b)).
Thus, we have  a non-zero imaginary part of the frequency for a single graphene
layer in a magnetic field. The non-zero imaginary part for
collective excitations for a 2D electron gas (2DEG) in semiconductors has been
established, out only for several layers of semiconductor \cite{Bakshi1,Bakshi2,Bakshi3,Bakshi4}.
See also  \cite{Bakshi1,Bakshi2,Bakshi3,Bakshi4}.)

The negative group velocity for $ql_H>1$ is caused by the magnetic field \cite{Chiu_Quinn}.
 We  used the same  parameters are employed in calculating   Fig.\
1, but we summed over a larger number of  Landau levels in the
conduction and valence bands.  Qualitatively, the results are the
same. The main differences are that the number of single-particle
excitation lines which are allowed is increased and the frequency of
the highest mode which is, of course, affected by the number of
Landau levels included in the sum. However,  the lower branches of
collective modes do not change significantly. Therefore, including
in the calculations the five lowest Landau levels in the conduction
and valence bands is justified.

\begin{figure}
\includegraphics[width = 2.0in] {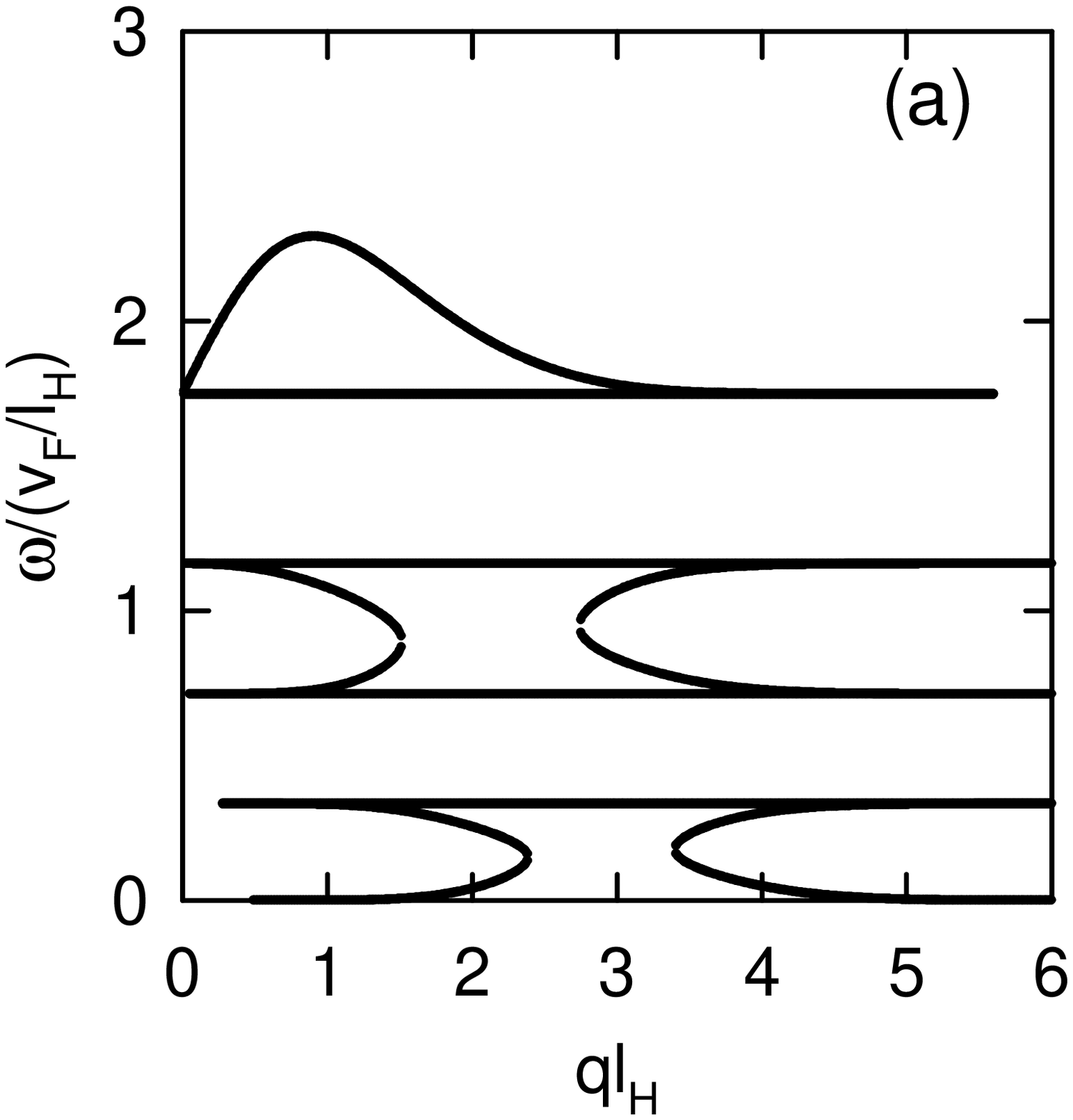}
\end{figure}
\begin{figure}
\includegraphics[width = 2.0in] {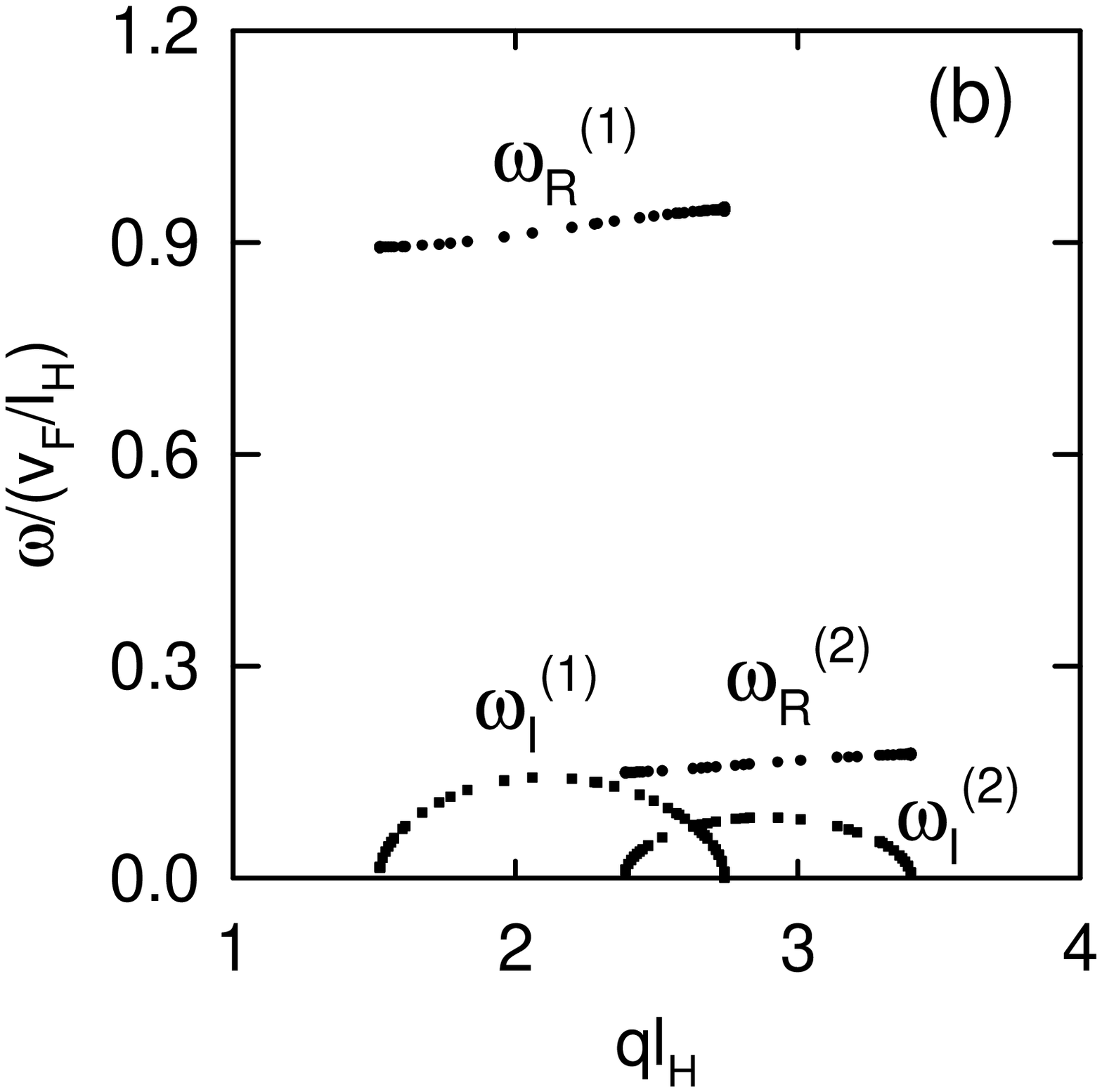}
\caption{Magnetoplasmon excitation energy as a
function of wave vector, in units of $l_H^{-1}$,
in a single graphene layer. (a) Real  frequency
solution. (b)  The real and imaginary parts of the
frequency satisfying the dispersion equation.}
\label{fsingle}
\end{figure}

\section{A graphite bilayer}
\label{bila}

While the electron effective mass in a graphene single layer is
 zero, a graphite bilayer consisting of a pair of parallel
graphite layers  with inter-planar separation $c/2$
 implies finite electron mass. The electron spectrum  a graphite
bilayer in a magnetic field  is very much different from the case of
a single graphene layer.  This is caused by interlayer hopping.
Here, $c/a = 2.802$ with $a =2.566 \AA$  denoting the lattice
constant \cite{Trickey}. The nearest-neighbor tight-binding
approximation yields a gapless state with parabolic bands touching
at the $K$ and $K^\prime$ points instead of conical bands
\cite{Novoselov_bilayer,Falko}. A graphite bilayer can be treated as
a gapless semiconductor. The eigenfunction
$\psi_{\alpha}(\mathbf{r})$ of an electron in a graphite bilayer in
a perpendicular magnetic field is given  for low-lying energy
excitations by \cite{Falko}

\begin{eqnarray}
\label{efunction_b_2}
\psi_{\alpha}^{(b)}(x,y) = \frac{C_{n}^{(b)}}
{\sqrt{L_{y}}}e^{ik_y y}
\left( \begin{array}{c} \Phi_{n}(x+l_H ^{2}k_y ) \\
s(n) Q_{n}\Phi_{n-2}(x+l_H ^{2}k_y )
\end{array}\right) \ ,
\end{eqnarray}
where $\displaystyle{\alpha=\{k_y ,n,s(n)\}}$,
$C_n^{(b)} = 1$ when $n=0$ or $n=1$ and $C_n^{(b)} = 1/\sqrt 2$
when $n\geq 2$. Also,  $Q_n = 0$ when $n=0$ or $n=1$ and
$Q_n = 1$ when $n\geq 2$
and $\Phi_n(x)$ is defined above.
The corresponding eigenenergy is
 $\epsilon_\alpha^{(b)} = s(n) \epsilon_n^{(b)} =
 s(n) \hbar\omega_{c}\sqrt{n(n-1)}$,
where $\omega_c=eB/m$ with
$m=\gamma_{1}/(2v^{2})$, $\gamma_1 = 0.39 $ eV and $v = 8\times 10^{5}m/s$
\cite{Falko} (compare to the electron spectrum in magnetic field in a
single graphene layer presented above).

\begin{figure}
\includegraphics[width = 2.0in] {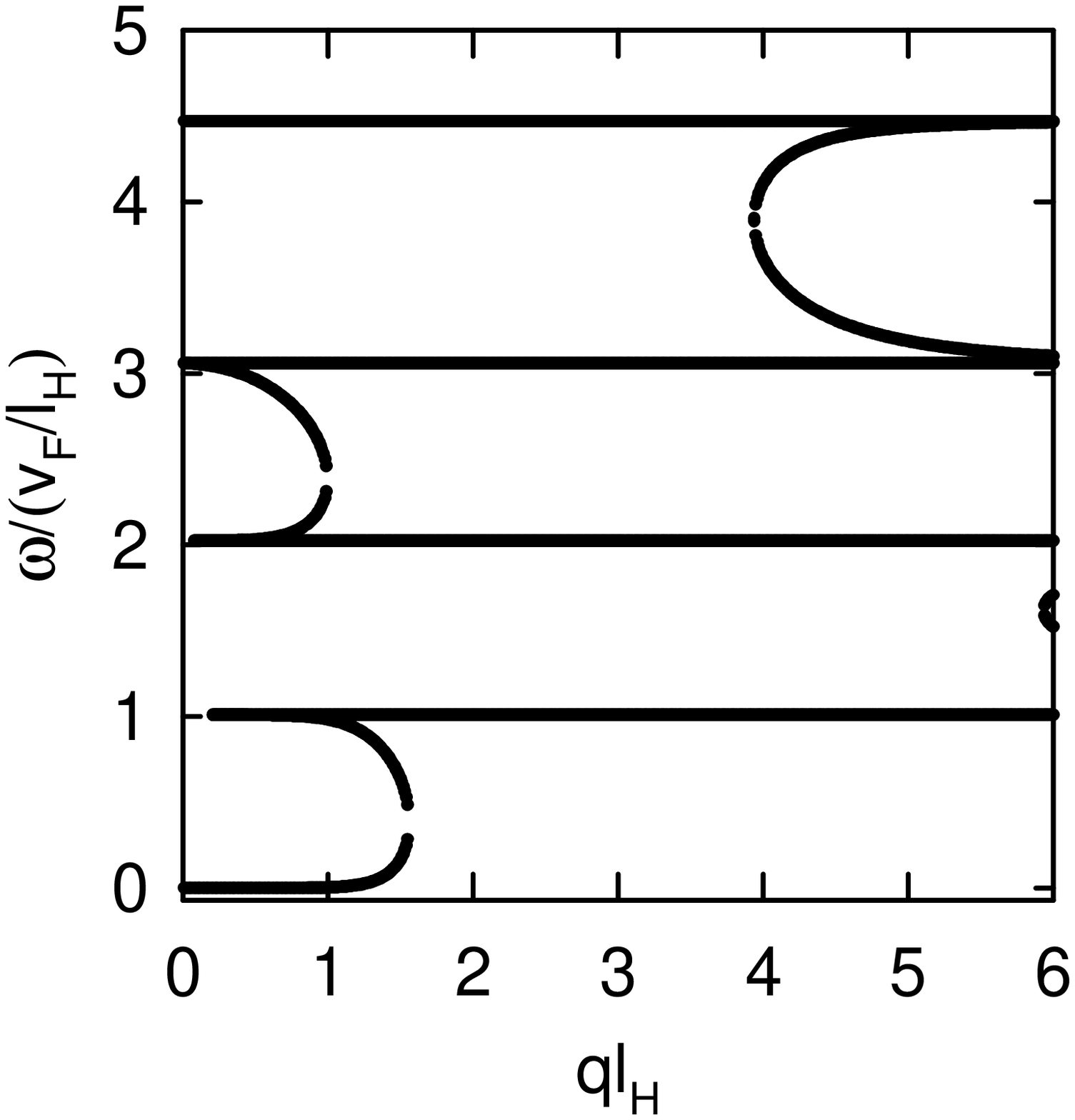}
\caption{Magnetoplasmon dispersion relation in a graphite
bilayer at  filling $\nu=1$ and $D/l_H = 0.1$. }
\label{fgraphite}
\end{figure}

 Following the procedure described above, it can be
seen that for the RPA dielectric function   $\varepsilon^{(b)}(q,\omega)$ for
a graphite bilayer, we must replace  the polarization by $\Pi^{(b)}(q,\omega)$
instead of $\Pi(q,\omega)$.
This is obtained from Eq.\  (\ref{pol}) by means of the eigenspectrum
$\epsilon_{n,s(n)}^{(b)}$
instead of $\epsilon_{n,s(n)}$. In this case, the form factor
$F_{ss^\prime}(n,n^\prime,q)$ is replaced by

\begin{eqnarray}
\label{Fkk_b}
&& F_{ss^\prime }^{(b)}(n,n^\prime,q)= A\left(C_{n}^{(b}C_{n^\prime}^{(b)}\right)^{2}
\left(\left|\int_{-\infty}^\infty dx\ \exp[iq_{x}x] \right. \right. \nonumber \\ &\times & \left. \left.
\Phi_{n}(x)\Phi_{n^\prime}(x+l_H ^{2}q_{y}) \right|^{2} \right. \nonumber \\
&+& \left. \left|s(n)s^\prime(n^\prime)Q_{n}Q_{n^\prime}\int_{-\infty}^\infty dx\
\exp[iq_{x}x] \Phi_{n-2}(x) \right. \right.  \nonumber \\ &\times & \left. \left. \Phi_{n^\prime-2}(x+l_H ^{2}q_{y})\right|^{2}\right)\ .
\end{eqnarray}
The eigenfunction of  bilayer Bernal graphene presented in
Eq.~(\ref{efunction_b_2}) was  obtained in Ref.\ [\onlinecite{Falko}]
by taking into account the
interlayer Coulomb interactions whose dominant contributions
are included. These are due to nearest-neighbor intralayer hopping
(see Fig.\ 1 in Ref.\ \onlinecite{Falko}). The interlayer Coulomb
interactions are included in Eq.\ (\ref{Fkk_b}) since they are contained in
the wavefunctions of Eq.\ (\ref{efunction_b_2}),  entering into
 the overlap integral  Eq.\ (\ref{Fkk_b}).
In Fig.\  \ref{fgraphite}, we present the dispersion relation.
 The four straight lines correspond to the single-electron transitions between different Landau levels. Three curved lines are the undamped
 magnetoplasmon excitations. For a range of wave vectors   the group velocity is negative due to the magnetic field, a result similar to those  in a single
graphene  layer. The transfer of energy between collective excitations and electrons appears only when the charged
particle velocity has the same value as the phase velocity of the
collective mode.

\section{A bilayer graphene}
\label{bilag}

For bilayer graphene with layer separation, $D$ and no interlayer
hoping, we have   the dispersion equation
\cite{Hwang_DasSarma,Eguiluz_Quinn,Madhukar,ggumbs}

\begin{eqnarray}
&& \sinh^{2}(qD)\left(2V_{c}(q)\Pi_{11}(q,\omega) -
\frac{\varepsilon_{1}}{\varepsilon_b} - \coth(qD)\right) \nonumber \\
&& \times \left(2V_{c}(q)\Pi_{22}(q,\omega) -
\frac{\varepsilon_{2}}{\varepsilon_{b}} -
 \coth(qD)\right)= 1\ ,
\label{diel_bil1}
\end{eqnarray}
where $\Pi_{jj}(q,\omega)$ is the polarization function of the charge
carriers on the first $j=1$ or the second $j=2$ graphene layer defined by
Eq.\  (\ref{pol}). We observe that in the limit $qD  \gg 1$,
Eq.\  (\ref{diel_bil1}) reduces to
the dispersion equation for magnetoplasmons in a single graphene layer.
If $\Pi_{11}(q,\omega) = \Pi_{22}(q,\omega) = \Pi(q,\omega)$, and $\varepsilon_{1}= \varepsilon_{2} = \varepsilon_{b}$, then we get
from Eq.~(\ref{diel_bil1}):

\begin{eqnarray}
&& \left[(2V_{c}(q)\Pi(q,\omega) - 1)
\left(1-e^{-2qD}\right) - \left(1 + e^{-2qD}\right)\right]
\nonumber \\
&& =\pm  2 e^{-qD} \  .
\label{diel_bil_simple}
\end{eqnarray}

\begin{figure}
\includegraphics[width = 2.0in] {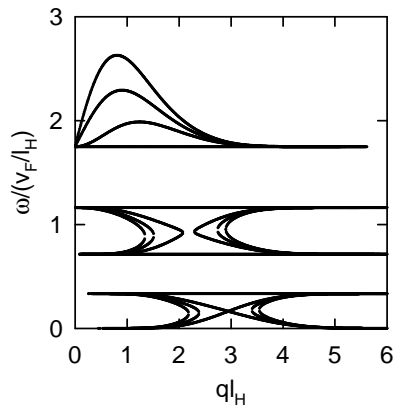}
\caption{Magnetoplasmon dispersion relation for bilayer graphene
 at $T=0$ K   with separation $D =  l_H $. Only the highest Landau level in
the valence is occupied and completely full.}
\label{fbilayer}
\end{figure}
The dispersion relation for magnetoplasmons in bilayer  graphene is
presented in Fig.\  \ref{fbilayer}. These results show that each
originally degenerate magnetoplasmon mode in each layer of an
isolated single graphene layer is shifted from their value by the
interlayer Coulomb interaction. For a range of wave vectors   the
group velocity is negative due to the magnetic field,  analogous to
the magnetoplasmon modes in a single graphene  layer.  A region of
instability also exists for bilayer graphene.

\section{An infinite periodic graphene superlattice}
\label{inf}

Let us consider an infinite periodic graphene superlattice consisting of
2D layers parallel to the $xy$-plane and located at $z=ld$ where
$l=0,\pm1,\pm2,\cdots, \pm\infty$  and $d$ is the period. The layers are
embedded in a medium with background dielectric constant $\varepsilon_b$.
The dispersion equation may be calculated in RPA in the same way described
in Refs.\  [\onlinecite{Das_Sarma_Quinn,Das_Sarma_prb_1983}].
It may be shown that the dispersion relation for magnetoplasmons
in a superlattice
is obtained by  solving $1 - V_{c}(q)\Pi(q,\omega)S(q,k_{z})=0$,
 where $\Pi (q,\omega)$ is the polarization function for
a single graphene layer
 defined by  Eqs.\  (\ref{pol})-\  (\ref{Fkk13}). Also,
$S(q,k_{z})$ is the structure factor
determining the phase coherence of the collective excitations in
different layers given by $S(q,k_{z})=
\sinh(qd)/(\cosh(qd)-\cos(k_{z}d)$. Note that the periodicity
ensures that $S(q,k_{z})$ is independent of the layer index $l$.
Also, the effective-mass model is employed to represent the low-frequency
electron  band structure of the layered graphene. At small
separations, the low-energy bands may be modified by the interlayer
atomic interactions.  In this case, the Landau levels may be
dispersive in the $k_z$ wave vector.

\begin{figure}
\includegraphics[width = 2.0in] {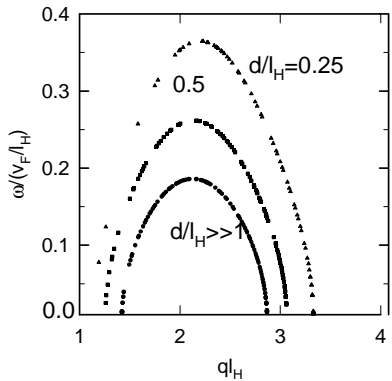}
\caption{The imaginary part of the magnetoplasmon energy
   in a single graphene layer compared to the results for
an  infinite superlattice  of graphene layers  at  $\nu=1$,
$k_{z}l_H = 0.1 $, for superlattice period $d =  0.25 l_H $,
$d =  0.5 l_H $, and $d \gg l_H $. }
\label{finfinite}
\end{figure}

 We have solved the magnetoplasmon dispersion equation in the complex
frequency plane. We present only the imaginary part of the solution
for various values of $d/l_H$ in Fig.\
\ref{finfinite} for $k_{z}l_H=0.1 $.

The results of our numerical calculations for an  infinite graphene
superlattice show that there are magnetoplasmon modes and modes
independent of the wave vector corresponding to single-particle transitions
between Landau levels. Due to a magnetic field, the group velocity is negative, as seen over a given range of wave vectors.
 This is  analogous to the magnetoplasmons in a single
graphene  layer. Energy transfer   from a charged particle to the
collective modes  occurs only when the charged particle's velocity has the same value as the phase velocity of the collective mode.
The most energetic collective mode increases for small wave vectors $q l_H < 1$  and is Landau damped for large wave vectors.
It is shown in Fig.\  \ref{finfinite}, that the imaginary part of the collective mode frequencies responsible for the  magnetoplasmon
instability is appreciably enhanced in an infinite
superlattice  of graphene layers compared to the single layer. Both the real and imaginary parts of
magnetoplasmon  frequencies are much larger in a graphene superlattice than in a single
graphene layer due to the superposition of the collective modes  corresponding to oscillations
from different layers occurring  in-phase. The amplification of the collective mode frequencies
increases when $a/l_H$ decreases.  According to Fig.\  \ref{finfinite}, it is clearly shown  that the
amplification of magnetoplasmons is increased as $a/l_H$ is reduced.  When  $a/l_H = 0.25$
  the magnetoplasmon frequencies are about twice as large compare to these at $a/l_H  \to\infty$. For $a/l_H = 0.5$,
  these corresponding frequencies are larger by a factor of $1.5$ relative to  the result when  $a/l_H \to \infty$.

While the 2D energy band is not suitable for describing the
low-frequency electronic properties of  bulk graphite
\cite{Charlier}, the calculated magnetoplasmon  frequencies obtained
from our superlattice model are valid. This is the case because  the
separation between  neighboring graphene layers in the superlattice
is much larger than in bulk graphite. In a graphene superlattice,
the distance between graphene layers can be sufficiently large,
e.g., as assumed in Fig. 4 $d=0.25 l_{H}$, $d=0.5 l_{H}$, or $d=
l_{H}$ (e.g., $ l_{H} = 66 {\AA}$ at $B=15 \mathrm{T}$), which is
large  compared to  the distance between carbon layers in  bulk
graphite which is $c/2$, where $c/a = 2.802$ with $a =2.566 \AA$
denoting the lattice constant.  The significance of this enhanced
magnetoplasmon instability in superlattices of graphene
for device applications may lie  in  possibly utilizing the energy
of the  amplified plasma  modes for electromagnetic radiation in
the THz regime,  leading to a potential new source of radiation
based on superlattices of graphene layers. For example, with an
applied  magnetic field $B= 10 T$, corresponding to filling
factor $\nu =1$, the magnetoplasmon  frequency is about $3.6 THz$.
Moreover, the advantage of    such sources of THz radiation is the
 fact that the frequencies corresponding to magnetoplasmon
instability leading to THz electromagnetic radiation decrease
when applied magnetic field increases and the parameter $a/l_H$
decreases which results in the possibilities of controlling  the
THz radiation frequencies by    changing the applied magnetic field.

\section{Discussion}
\label{disk}

We emphasize the appearance of a magnetoplasmon instability in a single
graphene layer even without  the application of an in-plane
current driving the charge carriers. This instability corresponds
to the finite imaginary part in the frequency of the collective
excitations in Fig.~\ref{fsingle}. There is a plasmon instability in
a bilayer semiconductor without an-in-plane current appears only in
a very small region (compared to the Fermi wave vector)
of the wave vector \cite{balassis}. This
difference in the spectrum of collective excitations in  graphene
structures compared  to layered semiconductors  is caused by the
screening properties of the dielectric function in
graphene \cite{Hwang_DasSarma,TAndo} and 2D semiconductors
\cite{Chiu_Quinn}.

\noindent
Acknowledgments: This work is supported by contract  \# FA9453-07-C-0207
of AFRL.

\end{document}